\documentclass[a4paper,12pt]{article}

\pdfoutput=1

\usepackage{amsmath}
\usepackage{amssymb}
\usepackage{amsfonts}
\usepackage{mathrsfs}
\usepackage{bbm}
\usepackage{graphicx,subfigure,booktabs,adjustbox}
\usepackage[numbers,sort&compress]{natbib}
\usepackage{verbatim}
\usepackage{float}
\usepackage{setspace}
\usepackage[colorlinks,
            linkcolor=blue,
            filecolor=blue,
            anchorcolor=blue,
            urlcolor=blue,
            citecolor=blue
            ]{hyperref}

\usepackage{color}
\usepackage{ulem}
\usepackage{bookmark}
\usepackage[utf8]{inputenc}

\numberwithin{equation}{section}

\newlength{\dinwidth}
\newlength{\dinmargin}
\makeatletter
\newcommand{\thickhline}{%
    \noalign {\ifnum 0=`}\fi \hrule height 1pt
    \futurelet \reserved@a \@xhline
}
\makeatother

\allowdisplaybreaks

\begin{document}

\title{\bf \Large \boldmath{$R_{D^{(\ast)}}$}, \boldmath{$R_{K^{(\ast)}}$} and neutrino mass in the 2HDM-III with right-handed neutrinos}

\author{Shao-Ping Li\footnote{ShowpingLee@mails.ccnu.edu.cn}, 
        Xin-Qiang Li\footnote{xqli@mail.ccnu.edu.cn},
        Ya-Dong Yang\footnote{yangyd@mail.ccnu.edu.cn}\, and
        Xin Zhang\footnote{xinzhang@mail.ccnu.edu.cn}\\[15pt]
{\small Institute of Particle Physics and Key Laboratory of Quark and Lepton Physics~(MOE),}\\[-0.3cm]
{\small Central China Normal University, Wuhan, Hubei 430079,
China}
}

\date{}
\maketitle
\vspace{0.2cm}

\begin{abstract}
\noindent Given that the two-Higgs-doublet model of type III (2HDM-III) has the potential to address the $R_{D^{(\ast)}}$ anomalies while the resolution to the $R_{K^{(\ast)}}$ deficits requires new degrees of freedom within this framework, we consider in this paper a unified scenario where the low-scale type-I seesaw mechanism is embedded into the 2HDM-III, so as to accommodate the $R_{D^{(\ast)}}$ and $R_{K^{(\ast)}}$ anomalies as well as the neutrino mass. We first revisit the $R_{D^{(\ast)}}$ anomalies and find that the current world-averaged results can be addressed at $2\sigma$ level without violating the bound from the branching ratio $\mathcal{B}(B_c^-\rightarrow \tau^- \bar{\nu})\leqslant 30\%$. The scenario predicts two sub-eV neutrino masses based on a decoupled heavy Majorana neutrino and two nearly degenerate Majorana neutrinos with mass around the electroweak scale. For the $R_{K^{(\ast)}}$ anomalies, the same scenario can generate the required Wilson coefficients in the direction $ C_{9 \mu}^{\rm NP}=-C_{10 \mu}^{\rm NP}<0$, with $\mathcal{O}(1)$ Yukawa couplings for the muon and the top quark.
\end{abstract}

\newpage

\section{Introduction}
\label{sec:intro}

Flavor physics is one of the most powerful probes of physics beyond the Standard Model (SM)~\cite{Buchalla:2008jp,Antonelli:2009ws}. Recently, several discrepancies between the SM predictions and the experimental measurements have been observed in $b\rightarrow c$ and $b\rightarrow s$ semi-leptonic transitions. The measured observables that can be used to test the lepton-flavor universality (LFU) are theoretically rather clean, because the involved hadronic uncertainties are cancelled to a large extent. Thus, the anomalies observed in these decays would suggest intriguing hints for LFU violating New Physics (NP) beyond the SM.

The LFU violating observables we first consider are the ratios $R_{D^{(\ast)}}$, which are defined as
\begin{align}
 R_{D^{(\ast)}}\equiv \frac{\mathcal{B}(\bar{B}\rightarrow D^{(\ast)} \tau \bar{\nu})}{\mathcal{B}(\bar{B}\rightarrow D^{(\ast)} \ell \bar{\nu})},
\end{align}
with $\ell=e$ or $\mu$, and have been measured by the BaBar~\cite{Lees:2012xj,Lees:2013uzd}, Belle~\cite{Huschle:2015rga,Sato:2016svk,Hirose:2016wfn,Hirose:2017dxl}, and LHCb~\cite{Aaij:2015yra,Aaij:2017uff,Aaij:2017deq} collaborations. The latest world-averaged results compiled by the Heavy Flavor Averaging Group (HFLAV)~\cite{HFLAV:2018} read: $R_{D^\ast}= 0.306\pm 0.013(\text{stat})\pm 0.007(\text{syst})$ and $R_D=0.407\pm 0.039(\text{stat})\pm 0.024(\text{syst})$, which indicate a combined deviation from the SM values $R^{\rm SM}_{D^\ast}\approx 0.26$ \cite{Fajfer:2012vx,Bernlochner:2017jka,Bigi:2017jbd,Jaiswal:2017rve} and $R^{\rm SM}_D\approx 0.30$ \cite{Lattice:2015rga,Na:2015kha,Bigi:2016mdz,Bernlochner:2017jka,Jaiswal:2017rve} at the level of $4\sigma$.
Thus far, feasible NP scenarios based on model-independent analyses~\cite{Fajfer:2012vx,Fajfer:2012jt,Becirevic:2012jf,Biancofiore:2013ki,Bhattacharya:2014wla,Alonso:2015sja,Bardhan:2016uhr,Colangelo:2016ymy,Bhattacharya:2016zcw,Ivanov:2017mrj,Alok:2017qsi,Azatov:2018knx,Feruglio:2018fxo,Colangelo:2018cnj} as well as model-dependent constructions such as leptoquarks~\cite{Deppisch:2016qqd,Bauer:2015knc,Freytsis:2015qca,Fajfer:2015ycq,Becirevic:2018afm,Li:2016vvp,
Assad:2017iib,Alok:2017jaf,Blanke:2018sro,Altmannshofer:2017poe} and two-Higgs-doublet models (2HDM)~\cite{Celis:2012dk,Crivellin:2012ye,Freytsis:2015qca,Crivellin:2015hha,Iguro:2017ysu} have been extensively studied towards an explanation of the  $R_{D^{(\ast)}}$ anomalies. In particular, the general 2HDM of type-III (2HDM-III) with tree-level flavor-changing neutral current (FCNC) can address the $R_{D^{(\ast)}}$ anomalies~\cite{Crivellin:2012ye,Crivellin:2015hha,Iguro:2017ysu}, but suffers severe constraint from the $B_c^-$ lifetime~\cite{Alonso:2016oyd,Li:2016vvp,Celis:2016azn,Akeroyd:2017mhr}.

On the other hand, the LFU violating observables $R_{K^{(\ast)}}$, which are defined as
\begin{align}
R_{K^{(\ast)}}\equiv\frac{\mathcal{B}(\bar{B} \rightarrow K^{(\ast)} \mu^+ \mu^-)}{\mathcal{B}(\bar{B}\rightarrow K^{(\ast)} e^+ e^-)},
\end{align}
have also been reported by the LHCb collaboration, giving
$R_{K}=0.745^{+0.090}_{-0.074}(\text{stat})\pm 0.036(\text{syst})$ in  $1\leqslant q^2\leqslant6~{\rm GeV}^2$~\cite{Aaij:2014ora}, $R_{K^\ast}=
0.66^{+0.11}_{-0.07}{\rm (stat)} \pm 0.03{\rm (syst)}$ in $0.045\leqslant q^2 \leqslant1.1 ~{\rm GeV}^2$ and $0.69^{+0.11}_{-0.07}{\rm (stat)} \pm 0.05{\rm (syst)}$ in $1.1\leqslant q^2 \leqslant6.0 ~{\rm GeV}^2$~\cite{Aaij:2017vbb},
where $q^2$ is the dilepton invariant mass squared. The $R_K$ result deviates from the SM value $R^{\rm SM}_K=1.00\pm0.01$~\cite{Hiller:2003js,Descotes-Genon:2015uva,Bordone:2016gaq}  in the same $q^2$ region at the level of $2.6\sigma$, while the $R_{K^\ast}$ measurements deviate from the SM predictions\footnote{The theoretical predictions for the ratio $R_{K^*}$ can be found in ref.~\cite{Aaij:2017vbb} and references therein.} by $2.1\sim2.3\sigma$ for the first and $2.4\sim2.5\sigma$ for the second $q^2$ region, depending on the theoretical predictions used~\cite{Aaij:2017vbb}. The $R_{K^{(*)}}$ deficits stir up both model-independent global analyses~\cite{Altmannshofer:2014rta,Ghosh:2014awa,Hurth:2014vma,Altmannshofer:2015sma,Descotes-Genon:2015uva,Geng:2017svp,Bardhan:2017xcc,Capdevila:2017bsm,DAmico:2017mtc,Ciuchini:2017mik,Ghosh:2017ber,Hurth:2017hxg,Alok:2017sui,Altmannshofer:2017yso,Hiller:2017bzc} and model-dependent NP constructions such as the $Z^\prime$ models~\cite{Sierra:2015fma,Buras:2013qja,Gauld:2013qja,Altmannshofer:2014cfa,Crivellin:2015mga,Crivellin:2015lwa,Celis:2015ara,Tang:2017gkz,Chiang:2017hlj,King:2017anf,Chala:2018igk,Guadagnoli:2018ojc} and the leptoquark models~\cite{Hiller:2014yaa,Gripaios:2014tna,Bauer:2015knc,Freytsis:2015qca,Fajfer:2015ycq,Becirevic:2016oho,Deppisch:2016qqd,Becirevic:2017jtw,Becirevic:2018afm}. It is generally found that reasonable explanations for the $R_{K^{(\ast)}}$ anomalies at the second $q^2$ region can be achieved, while the resolution to the $R_{K^\ast}$ deficit at the first $q^2$ region requires more involved NP scenario~\cite{Ghosh:2017ber,Alok:2017sui}. Therefore, we will not consider the latter in this paper. While the $R_{D^{(\ast)}}$ anomalies can be improved in the 2HDM-III with a particular up-quark Yukawa texture~\cite{Iguro:2017ysu}, the same scenario cannot address the $R_{K^{(\ast)}}$ deficits, because the resulting Wilson coefficients $C_{9,10}^{\rm 2HDM}$ (see eqs.~(50,51) in ref.~\cite{Iguro:2017ysu}) are universal for all lepton flavors. However, keeping further the electron and/or neutrino Yukawa couplings of both Higgs doublets in a general 2HDM-III can lead to lepton-flavor non-universal $C_{9,10}^{\rm 2HDM}$, and hence provide a viable resolution to the $R_{K^{(\ast)}}$ anomalies, as shown for example in ref.~\cite{Iguro:2018qzf}.

Besides the above two intriguing anomalies, there is another clear NP signature observed in neutrino oscillations that indicates nonzero neutrino masses~\cite{Mohapatra:2005wg}. The massive neutrinos, no matter how small their masses are, cannot be generated in the SM due to the absence of right-handed neutrino states as well as the requirement of renormalizability. In neutrino physics, there exist many interesting models that can address the neutrino mass problem, such as the type I-III seesaw models\footnote{We refer to the review~\cite{Mohapatra:2005wg} and references therein for these three different seesaw models.}, the inverse seesaw (ISS) model~\cite{Mohapatra:1986bd,Malinsky:2009gw,Dev:2012sg}, as well as the low-scale type-I seesaw (LSS-I) model~\cite{Pilaftsis:2005rv,Kersten:2007vk,Zhang:2009ac,Adhikari:2010yt,Ibarra:2010xw,Ibarra:2011xn}.
Given that the 2HDM-III considered in ref.~\cite{Iguro:2017ysu} has the potential to accommodate the $R_{D^{(\ast)}}$ anomalies, while the resolution to the $R_{K^{(\ast)}}$ deficits based on the same framework requires new degrees of freedom,  we will consider in this paper a unified scenario where the LSS-I mechanism is embedded into the 2HDM-III and discuss the compatibility of neutrino mass generation along with the explanation towards the $R_{K^{(\ast)}}$ deficits.

Our paper is organized as follows. We begin in Sec.~\ref{2HDM-III} with a brief overview of the 2HDM-III, and then revisit the $R_{D^{(\ast)}}$ anomalies, demonstrating that the current world-averaged results can be addressed at $2\sigma$ level without violating the bound  $\mathcal{B}(B_c^-\rightarrow \tau^-\bar{\nu})\leqslant 30\%$. In Sec.~\ref{2HDM-III+LSSI}, we combine the 2HDM-III with the LSS-I mechanism, and discuss the relevant neutrino mass problem and the lepton-flavor violating constraints from the processes $\ell_i\rightarrow \ell_j \gamma$.  In Sec.~\ref{sec:RK}, we determine the Wilson coefficients in the direction $ C_{9 \mu}^{\rm NP}=-C_{10 \mu}^{\rm NP}<0$, providing therefore an explanation for the $R_{K^{(\ast)}}$ deficits at $1\sigma$ level. Finally, our conclusions are made in Sec.~\ref{conclusion}.

\section{General 2HDM-III and \texorpdfstring{\boldmath{$R_{D^{(\ast)}}$}}{Lg} anomalies}
\label{2HDM-III}

\subsection{Framework of general 2HDM-III}

In the 2HDM~\cite{Gunion:1989we,Branco:2011iw}, an additional scalar doublet with hypercharge $+1$ is introduced to the SM field content. The most general scalar potential with a softly-broken $Z_2$ symmetry can be written as
\begin{align}
V&=m_1^2\Phi_1^{\dagger} \Phi_1 +m_2^2 \Phi_2^{\dagger} \Phi_2 -(m_{12}^2\Phi_1^{\dagger} \Phi_2 + {\rm H.c.})
+\dfrac{\lambda_1}{2}(\Phi_1^{\dagger} \Phi_1)^2+ \dfrac{\lambda_2}{2}(\Phi_2^{\dagger} \Phi_2)^2
\nonumber\\
&+\lambda_3\Phi_1^{\dagger} \Phi_1\Phi_2^{\dagger} \Phi_2
+\lambda_4\Phi_1^{\dagger} \Phi_2\Phi_2^{\dagger} \Phi_1+\left[ \dfrac{\lambda_5}{2}(\Phi_1^{\dagger} \Phi_2)^2+{\rm H.c.} \right].
\label{scalar-potential}
\end{align}
If CP conservation is imposed further on the potential, the parameters $m_{12}^2$ and $\lambda_5$ would be real. The two scalar doublets are usually parametrized as
\begin{equation}
\Phi_a=\left( \begin{array}{cc}  \varphi_a^+ \\ \dfrac{1}{\sqrt{2}}(v_a+\phi_a+i\chi_a)
\end{array}
\right),
\label{doublet parametrization}
\end{equation}
and the two vacuum expectation values satisfy $v=\sqrt{v_1^2+v_2^2}=246~\rm{GeV}$.
The physical mass eigenstates are obtained from rotations of the weak-interaction basis in the following way:
\begin{align}
\left( \begin{array}{cc} H \\ h
\end{array}
\right)&=\left( \begin{array}{cc} \cos \alpha & \sin \alpha \\ -\sin \alpha & \cos \alpha
\end{array}
\right) \left( \begin{array}{cc} \phi_1 \\ \phi_2
\end{array}
\right),\\[0.2cm]
\left( \begin{array}{cc} G(G^{\pm}) \\ A(H^{\pm})
\end{array}
\right)&=\left( \begin{array}{cc} \cos \beta & \sin \beta \\ -\sin \beta & \cos \beta
\end{array}
\right) \left( \begin{array}{cc} \chi_1(\varphi_1^{\pm}) \\ \chi_2(\varphi_2^{\pm})
\end{array}
\right),
\end{align}
with $\tan\beta=v_2/v_1$. Here $G$ and $G^{\pm}$ denote the Goldstone bosons, and $H^\pm$, $H(h)$ and $A$ are the physical charged, scalar and  pseudoscalar Higgs bosons, respectively.

The generic Yukawa Lagrangian in the 2HDM-III is given by
\begin{align}
-\mathcal L_{\rm int}=\overline{Q}_L(Y_1^u \tilde{\Phi}_1+Y_2^u \tilde{\Phi}_2) u_R+ \overline{Q}_L(Y_1^d \Phi_1+Y_2^d \Phi_2) d_R + \overline{E}_L(Y_1^\ell \Phi_1+Y_2^\ell \Phi_2) e_R + {\rm H.c.}.
\label{lag}
\end{align}
Here, $\tilde{\Phi}_i=i\tau_2 \Phi_i^\ast$ with $\tau_2$ being the Pauli matrix; $Q_L$ and $E_L$ denote the left-handed quark and lepton doublets, respectively; $u_R$, $d_R$ and $e_R$ are the right-handed singlets.
The physical eigenstates of fermions are obtained by performing the rotations
$f_{L,R}=V_{L,R}^f\,f_{L,R}^\prime$, where the primed fields denote the weak eigenstates. After transforming to the mass-eigenstate basis, the Lagrangian in eq.~\eqref{lag} gives rise to the tree-level scalar-mediated FCNCs.

A common way to parametrize these scalar-mediated FCNC effects is to define:
\begin{align}
X^f_i \equiv \dfrac{1}{\sqrt{2}}\,V_L^f\, Y_i^f\, V_R^{f\,\dagger},
\end{align}
where for $i=1$, $f=u$ and for $i=2$, $f=\ell,d$. A systematic analysis for the effective couplings $X^f_i$ has been given in ref.~\cite{Crivellin:2013wna}. It is found that all entries of $X^{d,\ell}_2$ are severely constrained by various flavor processes. For $X_1^{u}$, on the other hand, there are only tight constraints on the first two generations, while $\mathcal{O}(1)$ $X^u_{1,32}$ and $X^u_{1,33}$ are still allowed, which has also been found in refs.~\cite{Altunkaynak:2015twa,Iguro:2018qzf}. Based on these observations, we will show in the subsequent sections that $X^u_{1,32}$ and $X^u_{1,33}$ are crucial for accommodating the $R_{D^{(\ast)}}$ and $R_{K^{(\ast)}}$ anomalies, respectively.

\subsection{Revisiting the \texorpdfstring{\boldmath{$R_{D^{(\ast)}}$}}{Lg}  resolution in the 2HDM-III}
\label{RD(star)}

In the 2HDM-III, new scalar and pseudoscalar operators generated by the exchanges of charged Higgs bosons $H^\pm$ will contribute to the tree-level $b\rightarrow c \tau \bar{\nu}$ transitions\footnote{As the Wilson coefficients of these operators are proportional to the mass of the final-state lepton, we will assume that only the tauonic modes are affected significantly by these operators.}. The corresponding effective Hamiltonian is given by
\begin{align}
\mathcal{H}_{\rm eff}=\dfrac{4G_F V_{cb}}{\sqrt{2}}\left(C_{SL}\mathcal{O}_{SL}+C_{SR}\mathcal{O}_{SR}\right),
\end{align}
with
\begin{align}
\mathcal{O}_{SL}=\left(\bar{c}P_L b\right)\left(\bar{\tau}P_L\nu\right),\, \qquad\mathcal{O}_{SR}=\left(\bar{c}P_R b\right)\left(\bar{\tau}P_L\nu\right),
\end{align}
where $P_{R,L}=(1\pm \gamma_5)/2$ are the chiral projection operators.

Under the 2HDM-III, the ratios $R_{D}$ and $R_{D^{\ast}}$ can be expressed in terms of their SM counterparts, respectively, as
 \cite{Akeroyd:2003zr,Sakaki:2012ft,Fajfer:2012vx,Crivellin:2015hha}:
\begin{align}
 R_D&= R_D^{\rm SM}\left[1+1.5 ~\text{Re}\left(C_{SR}+C_{SL}\right)+1.0 \vert C_{SR}+C_{SL}\vert^2\right],
\nonumber \\[0.2cm]
R_{D^\ast}&=R_{D^\ast}^{\rm SM} \left[1+0.12 ~\text{Re}\left(C_{SR}-C_{SL}\right)+0.05\vert C_{SR}-C_{SL}\vert^2\right].
\label{RD form}
\end{align}
The pseudoscalar operator, with the corresponding coefficient $C_P=C_{SR}-C_{SL}$, contributes also to the purely leptonic decay $B_c^-\rightarrow \tau^- \bar{\nu}$, with the corresponding branching ratio given by
\begin{align}
\mathcal{B}(B_c^-\rightarrow \tau^- \bar{\nu})=\tau_{B_c}\,\dfrac{G_F^2 |V_{cb}|^2 m_{B_c} m_{\tau}^2f_{B_c}^2}{8\pi}\,\left(1-\dfrac{m_{\tau}^2}{m_{B_c}^2}\right)^2\,\left| 1+\dfrac{m_{B_c}^2}{(m_b+m_c) m_{\tau}}C_P\right|^2,
\label{Bc lifetime}
\end{align}
where $f_{B_c}$ is the $B^-_c$ decay constant, $\tau_{B_c}$ the $B_c^-$ lifetime, and $m_{b,c}$ the $\overline{\rm MS}$ quark masses. The constraint from the $B_c^-$ lifetime~\cite{Alonso:2016oyd,Li:2016vvp,Celis:2016azn,Akeroyd:2017mhr} requires $\mathcal{B}(B_c^-\rightarrow \tau^- \bar{\nu})\leqslant 30\%$\footnote{Here, to be more conservative, we do not adopt the more stringent constraint $\mathcal{B}(B_c^-\rightarrow \tau^- \bar{\nu})\leqslant 10\%$ obtained in ref.~\cite{Akeroyd:2017mhr}, because this bound depends on the widespread theoretical values used for $\mathcal{B}(B_c^-\rightarrow J/\psi \ell \bar{\nu})$.}, which is obtained as follows~\cite{Alonso:2016oyd,Celis:2016azn}: As the total width of the $B_c$ meson is distributed among modes induced by the partonic transitions $\bar{c}\to \bar{s}\bar{u}d~(47\%)$, $\bar{c}\to \bar{s}\ell\bar{\nu}~(17\%)$, $b\to c\bar{u}d~(16\%)$, $b\to c\ell\bar{\nu}~(8\%)$ and $b\to c\bar{c}s~(7\%)$~\cite{Beneke:1996xe}, one can infer that only $\leqslant 5\%$ of the experimentally measured width is attributed to the tauonic mode, including the scalar NP contribution. However, due to the sizable theory uncertainties in this estimate, $0.4~{\rm ps}\leqslant\tau_{B_c}\leqslant0.7~{\rm ps}$~\cite{Beneke:1996xe}, such a constraint can be relaxed up to a $\leqslant 30\%$ of the total width if the longer lifetime $\tau_{B_c}=0.7~{\rm ps}$ is taken as an input for the SM calculation, as suggested firstly in ref.~\cite{Alonso:2016oyd}. This results in the conservative bound $\mathcal{B}(B_c\rightarrow \tau\nu)\leqslant 30\%$, as is now commonly used in the literature.

Based on the allowed regions for the couplings $X^f_i$~\cite{Crivellin:2013wna}, a particular texture of $X^{u}_1$ was first considered in ref.~\cite{Crivellin:2015hha} to address the $R_{D^{(\ast)}}$ anomalies:
\begin{align}
X^u_1\equiv \dfrac{1}{\sqrt{2}}V_L^u Y_1^u V_R^{u \dagger}=\left(\begin{array}{ccc}
  0 & 0 &0 \\
  0 & 0 &0 \\
  0 & \epsilon_{tc} &\epsilon_{tt}
\end{array}\right).
\label{FCNC repre}
\end{align}
Recently, such a scenario is re-analyzed more thoroughly in ref.~\cite{Iguro:2017ysu}, concluding that it is possible (impossible) to accommodate the $1\sigma$ region of $R_{D^{(\ast)}}$ suggested by Belle (HFLAV)  under the constraint $\mathcal{B}(B_c^-\rightarrow \tau^- \bar{\nu})\leqslant 30\%$. However, we will show explicitly that, under the same constraint, the current world-averaged results for $R_{D^{(\ast)}}$~\cite{HFLAV:2018} could be addressed at $2\sigma$ level, based on the above $X^u_1$ texture.

At this point, it is interesting to mention that the measured differential distributions $d\Gamma(\bar{B}\rightarrow D^{(\ast)}\tau\bar{\nu})/dq^2$ by BaBar~\cite{Lees:2013uzd} and Belle~\cite{Huschle:2015rga,Abdesselam:2016cgx} can also provide complementary information to distinguish different NP models; see for example refs.~\cite{Celis:2016azn,Tanaka:2012nw}. However, as pointed out in refs.~\cite{Celis:2016azn,Iguro:2017ysu}, both of the two collaborations' results still have large uncertainties and rely on the theoretical models. Therefore, we will not consider these $q^2$ distributions as a further constraint throughout this paper.

To demonstrate that the current world-averaged $R_{D^{(\ast)}}$ results can be accommodated at $2\sigma$ level under the constraint from $\mathcal{B}(B_c^-\rightarrow \tau^- \bar{\nu})\leqslant30\%$, we calculate the relevant Wilson coefficients in a particular 2HDM-III framework where $Y_1^d=0$, $Y_2^\ell=0$ and the up-quark FCNC is determined by eq.~\eqref{FCNC repre}. In this case, only the coefficient $C_{SL}$ is significant in the large $\tan\beta$ regime, with its size being given by
 \begin{equation}
 C_{SL}(M_{H^\pm})\simeq\dfrac{V_{tb}}{V_{cb}}\dfrac{\tan\beta}{M_{H^\pm}^2} v\, m_\tau \,\epsilon_{tc},
 \end{equation}
evaluated at the NP scale $\mu_H=M_{H^\pm}$. Evolving it down to the $b$-quark mass scale, we get~\cite{Freytsis:2015qca}:
\begin{equation}
C_{SL}(m_b)=\left[\dfrac{\alpha_s(m_t)}{\alpha_s(m_b)}\right]^{-12/23} \left[\dfrac{\alpha_s(M_{H^\pm})}{\alpha_s(m_t)}\right]^{-4/7} C_{SL}(M_{H^\pm}).
\end{equation}
When considering the SM predictions for $R_{D^{(\ast)}}$, it should be pointed out that the soft-photon corrections to the decays $\bar{B}^0\rightarrow D^+\tau^-\bar{\nu}$ and $B^-\rightarrow D^0\tau^-\bar{\nu}$ relative to the ones with muon final state can lead to $4.4\%$ and $3.1\%$ enhancements in $R^{\rm SM}_{D^+}$ and $R^{\rm SM}_{D^0}$, respectively, which are larger than the current lattice-QCD uncertainty of $R_{D}^{\rm SM}$~\cite{deBoer:2018ipi}. Bearing this in mind, we will adopt therefore the $2\sigma$ ranges of the arithmetic averages for $R^{\rm SM}_{D^{{(\ast)}}}$ from ref.~\cite{HFLAV:2018} in our analysis.

\begin{figure}[t]
	\centering
	\includegraphics[width=0.46\textwidth]{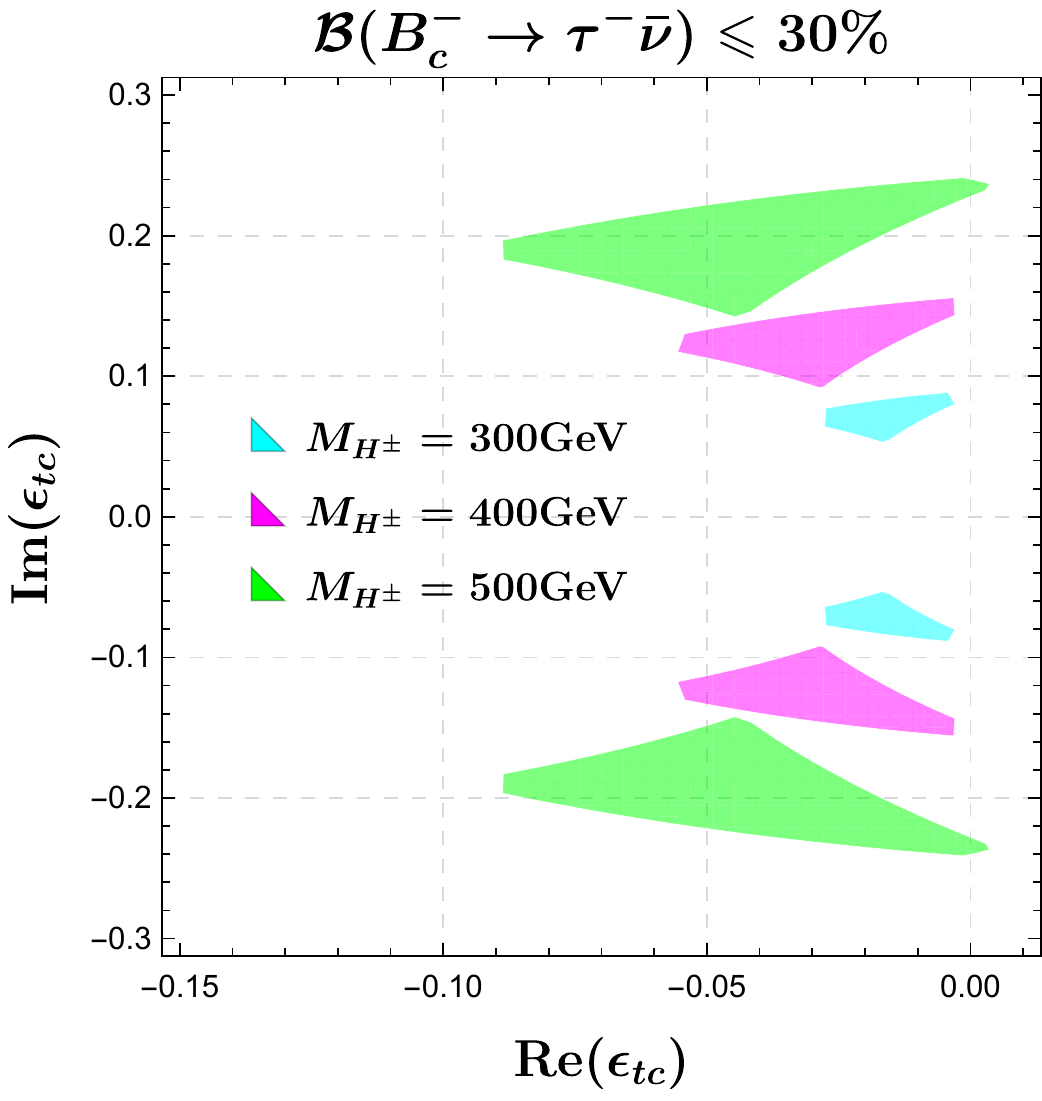}
	\caption{\small The allowed regions of $\epsilon_{tc}$ obtained from a $2\sigma$-level fit of the current world-averaged $R_{D^{{(\ast)}}}$ results, under the constraint $\mathcal{B}(B_c^-\rightarrow \tau^- \bar{\nu})\leqslant30\%$, with three different charged-Higgs boson masses and $\tan\beta=50$. }
	\label{BDtaufit}
\end{figure}

To fit the $2\sigma$ ranges of the current world-averaged $R_{D^{{(\ast)}}}$ results, we choose $\epsilon_{tc}$ as a free complex parameter and vary the charged-Higgs boson masses while fix $\tan\beta=50$. The SM parameters, if not stated otherwise, are taken from ref.~\cite{Patrignani:2016xqp} as follows: $G_F=1.166\times10^{-5}~{\rm GeV^{-2}}$, $\alpha_s(M_Z)=0.118$, $m_t=173.21~{\rm GeV}$, $m_b(m_b)=4.18~{\rm GeV}$, $m_c(m_c)=1.27~{\rm GeV}$, $m_\tau=1.78~{\rm GeV}$, $M_{B_c}=6.275~{\rm GeV}$, $\tau_{B_c}=0.507~{\rm ps}$, $f_{B_c}=0.434~{\rm GeV}$~\cite{Colquhoun:2015oha}, $|V_{cb}|=0.041$, and $|V_{tb}|=0.999$. The result is shown in Fig.~\ref{BDtaufit}. We can see that the constraint on $\epsilon_{tc}$ becomes more severe with smaller $M_{H^\pm}$. Note that a negative $\text{Re}(\epsilon_{tc})$ is required, because only $C_{SL}$ plays the significant role in the fit and the dominant contribution to $R_{D^{\ast}}$ comes from the interference term (see eq.~\eqref{RD form}). Generically, the magnitude $\vert\epsilon_{tc}\vert$ is bounded at $0.1-0.2$.

We conclude therefore that the $R_{D^{(\ast)}}$ anomalies can be addressed at $2\sigma$ level without violating the bound from $\mathcal{B}(B_c^-\rightarrow \tau^- \bar{\nu})\leqslant 30\%$ in the 2HDM-III. In the remaining sections, we will turn our attention to the neutrino mass as well as the $R_{K^{(\ast)}}$ anomalies in the same framework but with the LSS-I mechanism embedded into it.

\section{2HDM-III embedded with the LSS-I mechanism}
\label{2HDM-III+LSSI}

\subsection{Review of the LSS-I model}
\label{SM-LSSI}

The ISS~\cite{Mohapatra:1986bd,Malinsky:2009gw,Dev:2012sg} and LSS-I~\cite{Pilaftsis:2005rv,Kersten:2007vk,Zhang:2009ac,Adhikari:2010yt,Ibarra:2010xw,Ibarra:2011xn} models are the two popular candidates which allow for the low-scale heavy neutrino mass and the sizeable light-heavy neutrino mixing. In both of the two cases, the tiny neutrino mass is accounted for by a softly $U(1)$-symmetric breaking term. From the consideration of minimality, we will only discuss the LSS-I model, as the ISS model requires three more fermion singlets.

In the LSS-I model, at least two right-handed neutrino singlets should be introduced beyond the SM field content, to generate the phenomenologically viable pattern of neutrino masses. Such a minimal scenario with two right-handed neutrino singlets can be found \textit{e.g.} in  refs.~\cite{Zhang:2009ac,Liu:2017frs}. Here we will consider the three-generation case. The neutrino Yukawa interaction is now given by
\begin{align}
\overline{E}_L Y^{\nu}\tilde{H} N_R +\dfrac{1}{2}\overline{N}^c_R M_R N_R + {\rm H.c.},
\end{align}
where $H$ is the SM Higgs doublet and $N_R$ the right-handed neutrino singlet accompanied by a Majorana mass matrix $M_R$. After the spontaneous symmetry breaking, it leads to a full $6\times6$ neutrino mass matrix:
\begin{align}
-\dfrac{1}{2}\, \overline{n}_L \,M_{\nu} \, n_L^{c}+ {\rm H.c.}.
\end{align}
Here $n_L=\left(\nu_L,N_R^{c}\right)^T$. The mass matrix $M_\nu$ can be block-diagonalized by a $6\times6$ unitary matrix  $U^\nu$ defined in the following way~\cite{Ibarra:2010xw}:
\begin{align}
U^{\nu T} M_{\nu} \,U^\nu \equiv
\left(\begin{array}{cc} U^T_{\nu \nu} & U^T_{N\nu }\\ U^T_{\nu N} & U^T_{NN} \end{array} \right) \left(\begin{array}{cc} 0 & M_D \\ M_D^T & M_R \end{array} \right) \left(\begin{array}{cc} U_{\nu \nu} & U_{\nu N}\\U_{N \nu } & U_{NN} \end{array} \right) \simeq
\left(\begin{array}{cc} m_\nu & 0 \\0 & M_R \end{array} \right),
 \label{numass dia}
\end{align}
with the Dirac neutrino mass matrix $M_D= v Y^{\nu}/\sqrt{2}$. The light neutrino mass matrix $m_\nu \simeq -M_D M_R^{-1} M_D^T$ and the heavy Majorana neutrino mass matrix $M_R$ can be further diagonalized by the $3\times 3$ unitary matrices $\tilde{U}_P$ and $V_R$, respectively; \textit{i.e.},
\begin{align}
\hat{m}_\nu\equiv \tilde{U}_P^\dagger \,m_{\nu}  \,\tilde{U}_P^*,
\qquad
\hat{M}_N &\equiv V_R^\dagger \,M_R \,V_R^*,
\end{align}
where $\hat{m}_\nu \equiv \text{diag}(m_1,m_2,m_3)$ and $\hat{M}_N \equiv \text{diag} (M_1,M_2,M_3)$ denote the light and heavy neutrino mass eigenvalues, respectively.

As pointed out in ref.~\cite{Kersten:2007vk}, the tiny neutrino mass can be  induced by nearly degenerate heavy neutrinos with mass around TeV scale. Earlier in ref.~\cite{Pilaftsis:2005rv}, another scenario where three heavy neutrinos are nearly degenerate due to a softly $SO(3)$-symmetric breaking term was proposed to realize the electroweak-scale resonant leptogenesis and the small neutrino mass. In both of these two cases, however, the light-heavy neutrino mixing which is encoded in $U^*_{\nu N}\simeq M_D M_R^{-1}$~\cite{Ibarra:2010xw} cannot reach $\mathcal{O}(1)$ due to the indirect constraints from the low-energy precision data, such as the electroweak precision observables and the LFU tests~\cite{Antusch:2006vwa,Akhmedov:2013hec,Fernandez-Martinez:2015hxa,Fernandez-Martinez:2016lgt}. As a consequence, the severely restricted $U_{\nu N}$ cannot provide a solution to the $R_{K^{(\ast)}}$ anomalies via the neutrino-mediated box diagrams~\cite{He:2017osj}. Therefore, we have to introduce additional neutrino Yukawa interactions so as to provide an explanation for the $R_{K^{(*)}}$ deficits. In the next two subsections, we will illustrate that the additional neutrino Yukawa couplings can reach $\mathcal{O}(1)$ in the 2HDM-III framework.

\subsection{2HDM-III with electroweak-scale heavy neutrinos}

In the same spirit of ref.~\cite{Kersten:2007vk}, we consider the following  Yukawa Lagrangian added to eq.~\eqref{lag}:
\begin{align}
-\mathcal{L}_{N}=\overline{E}_L(Y_1^{\nu}\tilde{\Phi}_1+Y_2^{\nu} \tilde{\Phi}_2)N_R +\dfrac{1}{2}\overline{N}_R^c M_R N_R + {\rm H.c.}.
\label{nu Lag}
\end{align}
In the basis where the charged-lepton mass matrix is diagonal, we assume that the two Yukawa matrices $Y_{1,2}^\nu$ and the right-handed neutrino mass matrix $M_R$ have respectively the following textures:
\begin{align}
Y_1^\nu=\left(\begin{array}{ccc} x_1 & 0 & 0 \\ x_2 & 0 & 0 \\ x_3 & 0 & 0  \end{array} \right),\quad Y_2^\nu=\left(\begin{array}{ccc} 0 & 0 & y_1 \\ 0 & 0 & y_2 \\ 0 & 0 & y_3  \end{array} \right),\quad
M_R=\left(\begin{array}{ccc} 0 & M & 0 \\ M & \mu & 0 \\ 0 & 0 & M_3  \end{array} \right).
\label{matrix texture}
\end{align}
From the group-theoretical perspective, these textures manifest a global $U(1)$ symmetry under the charge assignments: $L(N_1)=-L(N_2)=1$, $L(N_3)=0$, $L(E_L)=1$, $L(\Phi_1)=0$  and $L(\Phi_2)=-1$. To avoid the scalar-mediated FCNC in the charged-lepton sector, we can assign to the right-handed charged leptons the $U(1)$ charges as: $L(e_R)=L(\mu_R)=L(\tau_R)=1$. On the other hand, we do not consider explicit $U(1)$ charge assignments for the quarks, because the explicit flavor-symmetry construction should now not only generate the needed FCNC texture given by eq.~\eqref{FCNC repre}, but also produce the already-known  pattern of the Cabibbo--Kobayashi--Maskawa mixing matrix~\cite{Cabibbo:1963yz,Kobayashi:1973fv}, which would become extremely nontrivial. Instead, we will assume that the Yukawa interactions in the quark sector are $U(1)$ invariant. In this case, the parameters $\mu$ in eq.~\eqref{matrix texture} and $m_{12}^2$ in eq.~\eqref{scalar-potential} (with $\lambda_5=0$) become the only sources to break softly the $U(1)$ symmetry.

The light neutrino mass matrix is now given by
\begin{align}
m_\nu \simeq -M_D M_R^{-1} M_D^T=\left(
\begin{array}{ccc}
 A& B & C \\
 B & D & E \\
 C & E & F\\
\end{array}
\right),
\label{mu-tau form}
\end{align}
with
\begin{align}
A&= \frac{v^2 x_1^2 \,\mu  \cos^2\beta}{2M^2}-\frac{v^2 y_1^2 \sin^2\beta}{2M_3},& 
B&=\frac{v^2 x_1 x_2 \,\mu
    \cos^2\beta}{2M^2}-\frac{v^2 y_1 y_2 \sin^2\beta}{2M_3},
    \nonumber \\
C&=\frac{v^2 x_1 x_3 \,\mu  \cos^2\beta}{2M^2}-\frac{v^2 y_1 y_3 \sin^2\beta}{2M_3},&
D&=\frac{v^2 x_2^2\,
   \mu \cos^2\beta}{2 M^2}-\frac{v^2 y_2^2 \sin^2\beta}{2 M_3},
   \nonumber \\
E&=\frac{v^2 x_2 x_3 \,\mu \cos^2\beta}{2
   M^2}-\frac{v^2 y_2 y_3 \sin^2\beta}{2 M_3},&
F&=\frac{v^2 x_3^2 \,\mu \cos^2\beta}{2M^2}-\frac{v^2 y_3^2 \sin^2\beta}{2M_3}.
\label{mu-tau form-2}
\end{align}
As the above neutrino mass matrix is of rank two, only two massive neutrinos are predicted in the considered scenario. Under the conditions that (i) $\tan\beta\gg 1$, (ii) the parameter $\mu$ is small, and (iii) $M_3\gg M\simeq \mathcal{O}(v)$, the sub-eV neutrino mass can be easily produced, without tuning the Yukawa couplings $x_i$ and $y_i$ to be extremely small. Explicitly, we find that the following set of parameters
\begin{align} \label{nuscale}
&x_i\sim\mathcal{O}(1),\quad
 y_i\sim\mathcal{O}(10^{-2}),\quad
 \tan\beta\sim\mathcal{O}(50),
\nonumber \\[0.2cm]
&M\sim\mathcal{O}(10^2)~\text{GeV},\quad
 M_3\sim\mathcal{O}(10^{10})~\text{GeV},\quad
 \mu\sim\mathcal{O}(10^{-7})~ \text{GeV},
\end{align}
would induce $m_\nu\sim 0.1$ eV\footnote{Realistic neutrino mass generation via the seesaw mechanism within the 2HDM framework has also been considered \textit{e.g.} in ref.~\cite{Campos:2017dgc}.}. Furthermore, the heavy Majorana neutrinos have mass eigenvalues $\hat{M}_N=\text{diag}(M-\mu/2,M+\mu/2,M_3)$. This indicates that the first two generations form a pseudo-Dirac neutrino~\cite{Bilenky:1978nj,Kersten:2007vk} with mass splitting proportional to $\mu$, while the third one is considered to decouple from the 2HDM-III field content when $M_3\gg M\simeq \mathcal{O}(v)$.

We now make remarks on the choice of the parameter set given by eq.~\eqref{nuscale}. The non-decoupled heavy neutrinos are assumed to reside at the electroweak scale, so that they can be produced directly at the high-energy colliders, providing therefore experimental tests for the LSS-I mechanism~\cite{Dev:2013wba,Das:2015toa,Khachatryan:2016olu,Das:2016hof,Das:2017zjc,Das:2017gke,Das:2017nvm,Das:2017rsu,Das:2018hph,Bhardwaj:2018lma}. One of the intriguing properties of the parameter $\mu$ in our case is that it is not necessary to be extremely small\footnote{In the ISS and LSS-I models, the scale of $\mu$ usually depends on the preference for the non-decoupled heavy neutrinos as well as the neutrino Yukawa couplings.}, because it is now accompanied by $\cos^2\beta$, the value of which is preferred to be small in light of the $R_{D^{(*)}}$ resolution within the 2HDM-III. Therefore, the hierarchy issue ($\mu\ll M$) can be relaxed to a large extent~\cite{Dias:2011sq,BhupalDev:2012zg}.

For the couplings $x_i$, as will be discussed in Sec.~4, an $\mathcal{O}(1)$ $x_2$ is required to address the $R_{K^{(*)}}$ anomalies. Such a muon-philic coupling also receives the indirect constraints studied in refs.~\cite{Antusch:2006vwa,Akhmedov:2013hec,Fernandez-Martinez:2015hxa,Fernandez-Martinez:2016lgt} for the light-heavy neutrino mixing parameters, but its contributions to the one-loop self-energy corrections of the $W/Z$ bosons were found to be negligible with electroweak-scale heavy neutrinos~\cite{Fernandez-Martinez:2015hxa}. Following the analysis made in ref.~\cite{Fernandez-Martinez:2016lgt}, we find that the contributions up to the one-loop order can be formally expressed as
\begin{equation}
\eta_{\ell} +  \frac{\vert x_i \vert^2}{16\pi^2}  \mathcal{S}_a(M, M_{H^\pm},M_{H,A}), 
\end{equation}
where $\eta_{\ell}$ represent the tree-level light-heavy neutrino mixing parameters, which are constrained to be of $\mathcal{O}(10^{-3})$~\cite{Fernandez-Martinez:2016lgt}, while $\mathcal{S}_a(M, M_{H^\pm},M_{H,A})$ denote the one-loop scalar functions. One can see that large $x_i$ may still be possible as their contributions are suppressed by the loop factor $1/(4\pi)^2$. At the same time, without any cancellations between the tree-level and one-loop contributions\footnote{If there exists cancellations to some extent, the constraints on $x_i$ can be further diluted. The cancellation scenario in which the light-heavy neutrino mixing parameters are allowed to be enhanced can be found \textit{e.g.} in refs.~\cite{Loinaz:2002ep,Loinaz:2004qc,Akhmedov:2013hec,Fernandez-Martinez:2015hxa}.}, we find that $\mathcal{S}_a(M, M_{H^\pm},M_{H,A})$ cannot exceed $\mathcal{O}(1)$ for $\vert x_i \vert\simeq \mathcal{O}(1)$, which can be readily satisfied with electroweak-scale neutrinos and Higgs bosons, say, $M\simeq \mathcal{O}(100~{\rm GeV})$ and $M_H\simeq M_A\simeq M_{H^\pm}\simeq\mathcal{O}(200~{\rm GeV})$.

As stressed in ref.~\cite{Fernandez-Martinez:2016lgt}, the lepton-flavor violating transitions $\ell_i\rightarrow \ell_j \gamma$ give one of the most severe constraints on the light-heavy neutrino mixing parameters. Thus, we will consider such constraints on the $x_i$ parameters with $x_2\simeq \mathcal{O}(1)$ in the next subsection. Specifically, we will analyze the process $\tau\rightarrow \mu \gamma$, while the more severe constraint from $\mu\rightarrow e \gamma$ that sets bound on the product $x_1 x_2$ can be simply avoided if $x_1\rightarrow 0$~\cite{Fernandez-Martinez:2016lgt,Lindner:2016bgg}.

\subsection{\texorpdfstring{\boldmath{$\tau\rightarrow \mu\gamma$}}{Lg} constraint}

In our scenario, the ratio between the decay width of $\tau\rightarrow\mu\gamma$ with respect to that of $\tau\rightarrow\mu\nu\bar{\nu}$ is given by
\begin{align}                                                \mathscr{B}(\tau\rightarrow\mu\gamma)&\equiv\dfrac{\Gamma(\tau\rightarrow \mu\gamma)}{\Gamma(\tau\rightarrow\mu\nu\bar{\nu})}
\nonumber \\
&=\dfrac{s_W^4}{384\pi^3 \alpha_{\rm em}}\,\dfrac{M_W^4}{M_{H^{\pm}}^4}\,\left|x_2 x_3 \right|^2\, \left[\dfrac{2\lambda^3+3\lambda^2-6\lambda^2\log(\lambda)-6\lambda+1}{(\lambda-1)^4}\right]^2,
\label{mutau data}
\end{align}
where $\lambda=M^2/M_{H^{\pm}}^2$, $s_W=\sin\theta_W$ with $\theta_W$ being the weak mixing angle, and $\alpha_{\rm em}$ is the fine-structure constant. In the above result, we have neglected the small Yukawa couplings in the charged-lepton part.

\begin{figure}[t]
	\centering
	\includegraphics[width=0.46\textwidth]{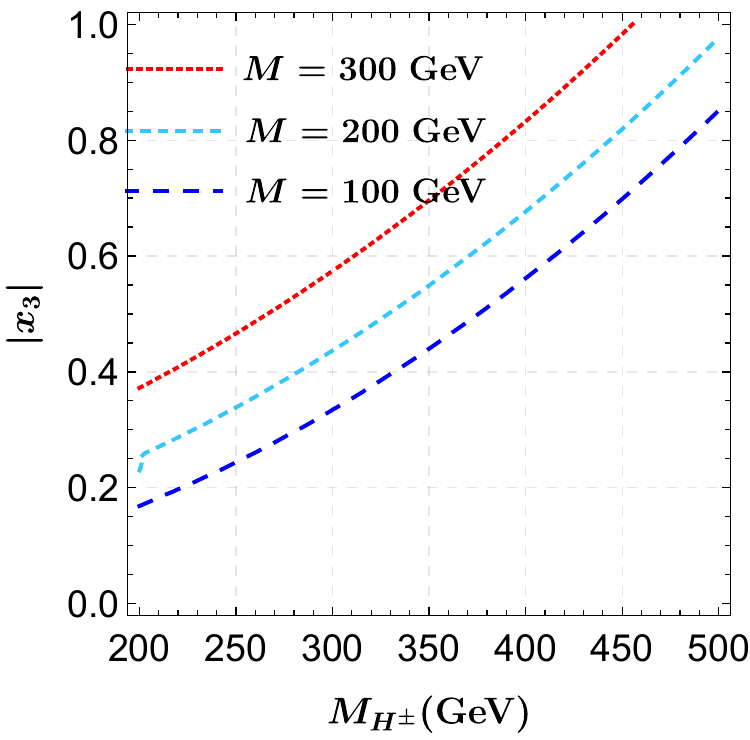}\quad
	\includegraphics[width=0.46\textwidth]{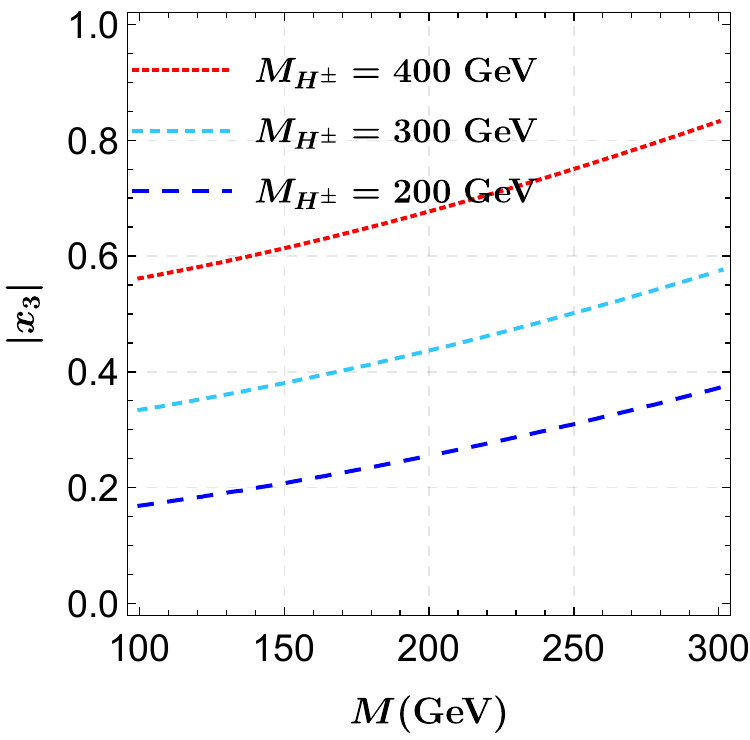}
	\caption{\small Constraint from the ratio $\mathscr{B}(\tau\rightarrow \mu \gamma)$ defined by eq.~\eqref{mutau data}. Left:  $(M_{H^\pm},\vert x_3\vert)$ contours with different heavy neutrino masses. Right: $(M,\vert x_3\vert)$ contours with different charged-Higgs boson masses. The regions below the curves are allowed by the current data.}
	\label{mutau plot}
\end{figure}

Fixing $x_2=1$, we show in Fig.~\ref{mutau plot} the contours in the $(M_{H^\pm},\vert x_3\vert)$~(left) and $(M,\vert x_3\vert)$~(right) planes, respectively. The regions below the curves are allowed by the experimental data with the inputs taken from ref.~\cite{Patrignani:2016xqp} as follows: $\sin^2\theta_W=0.2315$, $M_W=80.385~{\rm GeV}$, $\mathcal{B}(\tau\rightarrow \mu\gamma)\lesssim 4.4\times 10^{-8}$ and $\mathcal{B}(\tau \rightarrow\mu\nu\bar{\nu})=0.17$. We can see from Fig.~\ref{mutau plot} that $\vert x_3\vert$ is required to be small in order to comply with the $\tau\rightarrow \mu \gamma$ constraint. However, $\vert x_3\vert$ can still increase when $M$ or $M_{H^{\pm}}$ becomes larger.

Finally, we discuss the neutrino mixing parameters observed in the neutrino oscillation experiments. It was noticed that viable neutrino mixing pattern can be reproduced with $x_1=0$~\cite{Kersten:2007vk}. In this limit, the well-known tri-bimaximal mixing pattern (see \textit{e.g.} the review~\cite{Altarelli:2010gt}) with an inverted mass hierarchy $m_2>m_1>m_3=0$ can be obtained if $x_2=x_3$, $y_2=y_3$ and $D=(A+B)/2$ (see eqs.~\eqref{mu-tau form} and \eqref{mu-tau form-2}), which is motivated by the analysis made in ref.~\cite{Zhang:2009ac}. Certainly, the tri-bimaximal mixing pattern should be modified in order to generate nonzero reactor angle (see \textit{e.g.} the updated global fit for the neutrino oscillation data~\cite{Esteban:2016qun}), which, however, cannot be realized in the $x_1=x_3=0$ limit\footnote{We thank the referee for pointing out this unrealistic limit.}. For specific parameter choices, we refer to ref.~\cite{Kersten:2007vk} for details.

\section{\texorpdfstring{\boldmath{$R_{K^{(\ast)}}$}}{Lg} deficits in the 2HDM-III embedded with the LSS-I mechanism}
\label{sec:RK}

\subsection{Theoretical \texorpdfstring{\boldmath{$R_{K^{(\ast)}}$}}{Lg} explanation}
\label{RK anomaly solution}

In our analysis, we will focus only on the following subsets of operators which are directly responsible for the transition $b\rightarrow s \mu^+ \mu^-$~\cite{Buchalla:1995vs}:
\begin{align}
\mathcal{O}_7=&\dfrac{e}{16\pi^2} m_b \left(\bar{s} \sigma_{\mu\nu}P_R b\right)F^{\mu\nu},  & \mathcal{O}_7^\prime=&\dfrac{e}{16\pi^2} m_b \left(\bar{s} \sigma_{\mu\nu}P_L b\right)F^{\mu\nu},\\[0.2cm]
\mathcal{O}_9=&\dfrac{\alpha_{\rm em}}{4\pi} \left(\bar{s} \gamma_{\mu}P_L b\right)\left(\bar{\mu} \gamma^{\mu} \mu \right), & \mathcal{O}_9^{\prime}=&\dfrac{\alpha_{\rm em}}{4\pi} \left(\bar{s} \gamma_{\mu}P_{R} b\right)\left(\bar{\mu} \gamma^{\mu} \mu \right),\\[0.2cm]
\mathcal{O}_{10}=&\dfrac{\alpha_{\rm em}}{4\pi} \left(\bar{s} \gamma_{\mu} P_L b\right) \left(\bar{\mu} \gamma^{\mu} \gamma_5 \mu \right),& \mathcal{O}_{10}^\prime=&\dfrac{\alpha_{\rm em}}{4\pi} \left(\bar{s} \gamma_{\mu} P_R b\right) \left(\bar{\mu} \gamma^{\mu} \gamma_5 \mu \right).
\end{align}
Thus far, there are extensively model-independent analyses on the Wilson coefficients $C_{7,9,10}^{(\prime)}$ by fitting to the $R_{K^{(*)}}$ deficits as well as the various available data on $b\rightarrow s \ell^+ \ell^-$ and $b\rightarrow s \gamma$ transitions, such as the (differential) branching ratios $\mathcal{B}(B\rightarrow K^{(\ast)} \mu^+ \mu^-)$ and $\mathcal{B}(B_s\rightarrow \phi \mu^+ \mu^-)$, the (optimised) angular observables in $B^0\rightarrow K^{\ast0} \mu^+ \mu^-$ and $B_s\rightarrow \phi \mu^+ \mu^-$, and the branching ratio of the inclusive decay $B\rightarrow X_s \mu^+ \mu^-$~\cite{Altmannshofer:2014rta,Ghosh:2014awa,Hurth:2014vma,Altmannshofer:2015sma,Descotes-Genon:2015uva,Geng:2017svp,Capdevila:2017bsm,DAmico:2017mtc,Ciuchini:2017mik,Ghosh:2017ber,Hurth:2017hxg,Alok:2017sui,Altmannshofer:2017yso,Hiller:2017bzc}. It is consistently found that the NP in the muon sector is preferred, whereas no preference for the NP in the electron mode was favored~\cite{Altmannshofer:2014rta,Ghosh:2014awa,Hurth:2014vma,Altmannshofer:2015sma,Descotes-Genon:2015uva,Geng:2017svp,Capdevila:2017bsm,DAmico:2017mtc,Ciuchini:2017mik,Ghosh:2017ber,Hurth:2017hxg,Alok:2017sui,Altmannshofer:2017yso,Hiller:2017bzc}. Through the one-dimensional fits, it is found that the most preferred scenarios fall into the following three directions: (I) $C_{9 \mu}^{\rm NP}<0$, (II) $C_{9 \mu}^{\rm NP}=-C_{10 \mu}^{\rm NP}<0$, and (III) $C_{9 \mu}^{\rm NP}=-C_{9 \mu}^{\prime \rm NP}<0$. However, the scenario (III) predicts $R_K=1$ and hence cannot explain the  $R_{K^{(*)}}$ deficits simultaneously. In ref.~\cite{Alok:2017sui}, it is further found that the scenario (II) can provide a better fit in light of the LHCb measurement of $R_{K^{*}}$~\cite{Aaij:2017vbb}. Accordingly, we will investigate if this interesting scenario could be reproduced in our framework.

\begin{figure}[t]
	\centering
	\includegraphics[width=0.99\textwidth]{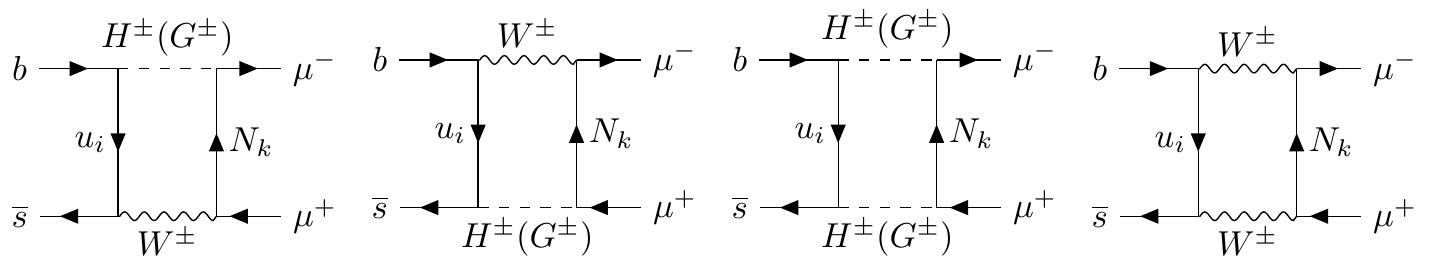}
	\caption{\small Box diagrams contributing to $b\rightarrow s \mu^+\mu^-$ transition in the 2HDM-III embedded with the LSS-I mechanism.}
	\label{9boxes}
\end{figure}

In our scenario, the Wilson coefficients $C_{7,9,10}^\prime$ will receive a suppression factor $1/\tan\beta$, which can be also seen from refs.~\cite{Crivellin:2013wna,Iguro:2017ysu}. Although a sizeable $C_7$ can be generated in our scenario, it is severely constrained by the inclusive decay $B\rightarrow X_s\gamma$\footnote{In ref.~\cite{Li:2018aov}, we have shown explicitly that $C_7$ can be significantly reduced due to a destructive cancellation if a nonzero $\epsilon_{ct}$ is introduced in eq.~\eqref{FCNC repre}, especially in the case for a relatively light charged Higgs boson.}. Hence only $C_{9,10}$ are relevant to our discussion for the $R_{K^{(*)}}$ anomalies. We find that the Feynman diagrams depicted in Fig.~\ref{9boxes} can give sizeable contributions to $C_{9 \mu}^{\rm NP}=-C_{10 \mu}^{\rm NP}<0$, which is favored by the scenario (II). The dominant contribution comes from the third diagram with two charged Higgs bosons running in the loop, because the vertex $H^{\pm} N \mu^\mp$ allows a sizeable coupling  ($\mathcal{O}(1)$) while the $W^{\pm}N \mu^\mp$ coupling is constrained to be $\mathcal{O}(10^{-2})$~\cite{Antusch:2006vwa,Akhmedov:2013hec,Fernandez-Martinez:2015hxa,Fernandez-Martinez:2016lgt}. To this end, for simplicity, we will consider only the contribution coming from this diagram.

After a direct calculation, the corresponding Wilson coefficients are given by
\begin{align}
C_{9 \mu}^{\rm NP}=-C_{10 \mu}^{\rm NP}
=\dfrac{v^4}{32 s_{W}^2 M_W^4}\,\sum_{i=c,t} \vert\epsilon_{ti}\vert^2\,\vert x_2\vert^2\,I(x,y,z_i),
\label{C9=-C10}
\end{align}
with $\epsilon_{ti}$ given by eq.~\eqref{FCNC repre}. The scalar function $I(x,y,z_i)$ is defined as
\begin{align}
I(x,y,z_i)&=\frac{y^2 \log \left(x/y\right)}{(x-y)^2 (y-z_i)}+\frac{z_i^2 \log
   \left(x/z_i\right)}{(x-z_i)^2 (z_i-y)}-\frac{x}{(x-y) (x-z_i)},
   \label{RK scalar form}
\end{align}
where $x=M_{H^{\pm}}^2/M_W^2$, $y=M^2/M_W^2$, and $z_i=m_i^2/M_W^2$. Here we have neglected the mass splitting between the two non-decoupled heavy Majorana neutrinos. The decoupled Majorana neutrino, on the other hand, does not play any role in the box diagrams because its couplings to the 2HDM fields are suppressed by the inverse of its mass.

Finally, we need to mention that there are also contributions from the $Z$- and $\gamma$-penguin diagrams, giving rise to the lepton-flavor universal Wilson coefficients $C_{9\ell}^{\rm NP}$ and $C_{10\ell}^{\rm NP}$, with $\ell=e$, $\mu$, or $\tau$. However, using the formulae given in  ref.~\cite{Iguro:2017ysu}, we have checked numerically that these contributions are small for $M_{H^{\pm}}\simeq 500$ GeV, $\vert\epsilon_{tc}\vert\leqslant 0.5$, and $\vert\epsilon_{tt}\vert\leqslant 1$. Hence we will not consider these contributions in the following numerical analysis.

\subsection{Numerical \texorpdfstring{\boldmath{$R_{K^{(\ast)}}$}}{Lg} analysis}
\label{RK data}

The free parameters in eq.~\eqref{C9=-C10} are $\epsilon_{tc,tt}$ and $x_2$, together with the heavy neutrino mass $M$ and the charged-Higgs boson mass $M_{H^\pm}$. However, as shown in Fig.~\ref{BDtaufit}, there exists a strong correlation between $\epsilon_{tc}$ and $M_{H^\pm}$ stemming from the $R_{D^{(\ast)}}$ fits. Therefore, we choose three typical values of $(\vert \epsilon_{tc}\vert,M_{H^\pm})$: $(0.08,300~\text{GeV})$, $(0.14,400~\text{GeV})$, and $(0.21,500~\text{GeV})$ with $\tan\beta=50$ in our numerical analysis.

\begin{figure}[t]
	\centering
	\includegraphics[width=0.46\textwidth]{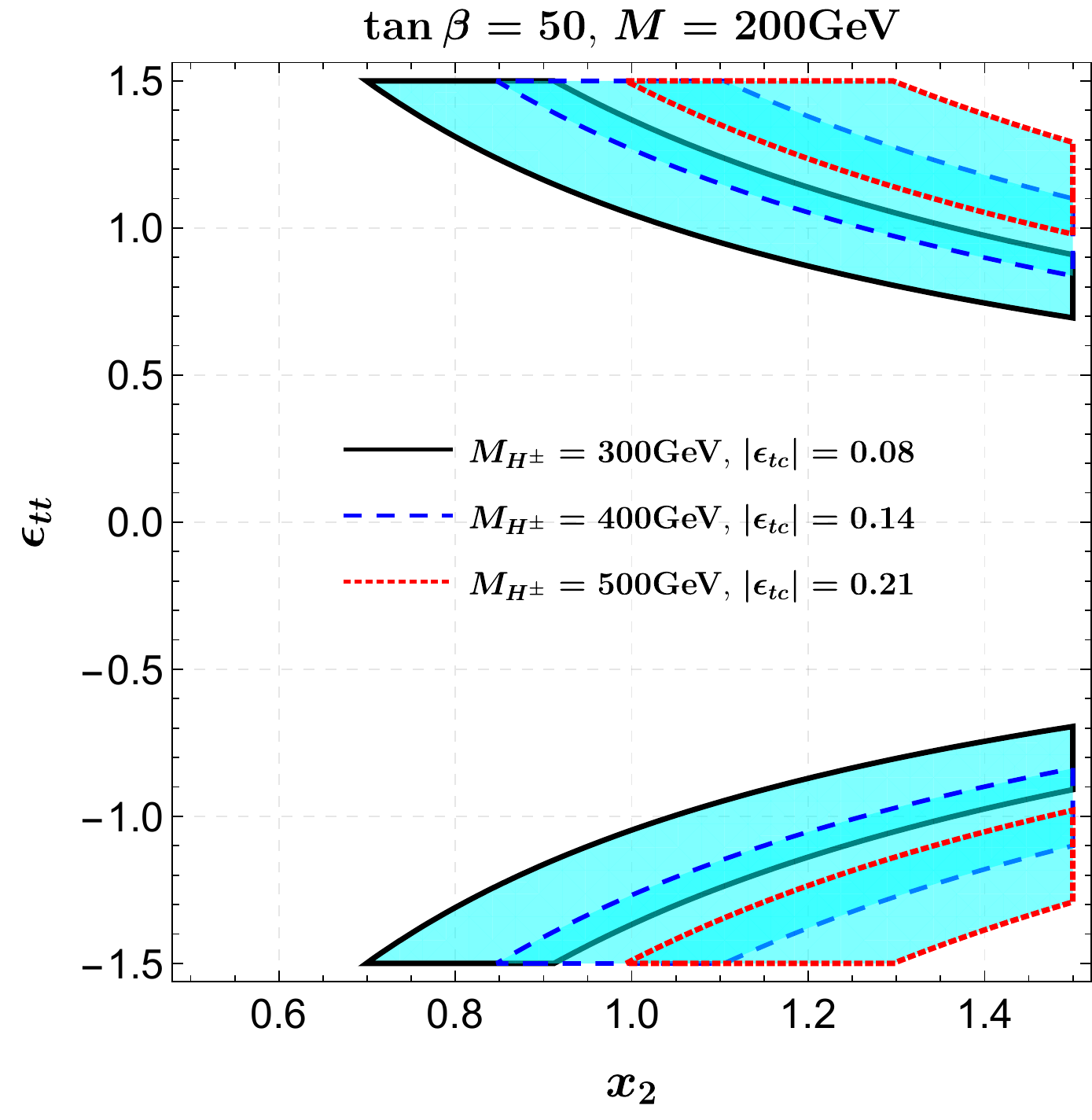}
	\caption{\small Constraint on the parameters $x_2$ and $\epsilon_{tt}$ using the $1\sigma$ range of the Wilson coefficients $C_{9 \mu}^{\rm NP}=-C_{10 \mu}^{\rm NP}<0$ obtained through a global fit to the $R_{K^{(*)}}$ deficits as well as the various available data on $b\rightarrow s \ell^+ \ell^-$ and $b\rightarrow s \gamma$ transitions~\cite{Altmannshofer:2017yso}.}
	\label{RKsol}
\end{figure}

In Fig.~\ref{RKsol}, we plot the $(x_2, \epsilon_{tt})$ plane (assuming $x_2>0$) by using the 1$\sigma$ range of the Wilson coefficients $C_{9 \mu}^{\rm NP}=-C_{10 \mu}^{\rm NP}<0$ obtained through a global fit to the $R_{K^{(*)}}$ deficits as well as the various available data on $b\rightarrow s \ell^+ \ell^-$ and $b\rightarrow s \gamma$ transitions. Here we have fixed $M=200~{\rm GeV}$ as the scalar function (eq.~\eqref{RK scalar form}) is insensitive to the neutrino mass around the electroweak scale. As can be seen from Fig.~\ref{RKsol}, $\mathcal{O}(1)$ $x_2$ and $\vert\epsilon_{tt}\vert$ are required to account for the $R_{K^{(\ast)}}$ deficits. When the other eight box diagrams depicted in Fig.~\ref{9boxes} are also taken into account with a sizeable $W^\pm N \mu^\mp$ coupling~\cite{Fernandez-Martinez:2016lgt}, the required sizes of $x_2$ and  $\vert\epsilon_{tt}\vert$ can both be reduced. However, these contributions are not explicitly taken into account when making the plots in Fig.~\ref{RKsol}, because in this case more parameters would be involved.

It should be pointed out that the parameters $\epsilon_{tt}$ and $M_{H^\pm}$ are also tightly constrained by the $B_s-\bar{B}_{s}$ mixing and the $b\rightarrow s\gamma$ transitions, with the findings that $\epsilon_{tt}\lesssim1$ for $M_{H^\pm}\lesssim500$ GeV~\cite{Mahmoudi:2009zx,Crivellin:2013wna}, which are compatible with the ones required for explaining the $R_{K^{(\ast)}}$ deficits. Thus, our scenario can provide an explanation for the $R_{D^{(\ast)}}$ and $R_{K^{(\ast)}}$ anomalies, while complying with these tight constraints. On the other hand, the $\mathcal{O}(1)$ coupling $x_2$, besides its contribution to $R_{K^{(\ast)}}$, also contributes to the muon $g-2$ dominantly at the one-loop level. However, this contribution is only of $\mathcal{O}(10^{-10})$ for $M_{H^\pm}\gtrsim 100$ GeV~\cite{Ma:2001mr}, which is smaller than the current experimental data~\cite{Patrignani:2016xqp} by an order of magnitude. It is therefore difficult to provide a resolution to the muon $g-2$ excess in the same scenario. In a follow-on paper~\cite{Li:2018aov}, we will show that large contributions to the muon $g-2$ can come from the two-loop Barr-Zee type diagrams. If the muon $g-2$ excess is attributed to these two-loop Barr-Zee contributions, large $\epsilon_{tt}$ and relatively light charged Higgs boson would be required. In this case, the constraints from $B_s-\bar{B}_{s}$ mixing and $b\rightarrow s \gamma$ transitions would become very severe. However, with a nonzero $\epsilon_{ct}$ introduced to $X_1^u$ (see eq.~\eqref{FCNC repre})~\cite{Crivellin:2013wna,Altunkaynak:2015twa,Iguro:2018qzf,Chen:2018hqy}, the muon $g-2$ anomaly can still be addressed while the constraints from these processes are satisfied at the same time~\cite{Li:2018aov}.

Finally, it should be mentioned that, due to the presence of $\mathcal{O}(1)$ parameters $x_2$ and $\epsilon_{tt}$, the decay modes $H^+ \to t b$ and $H^+ \to \mu N$ can have large branching ratios, depending on the explicit mass spectrum of heavy neutrino, top quark and charged Higgs boson. For the $H^+ \to t b$ decay, a recent search performed at the LHC has put upper limits on the cross section times branching ratio $\sigma(pp\rightarrow t b H^+)\times \mathcal{B}(H^+ \rightarrow t b)$ for $M_{H^\pm}=200-2000$ GeV~\cite{Aaboud:2018cwk}. As for the $H^+ \to \mu N$ decay, the detection of the final states relies on the decay products of the heavy neutrinos and hence would involve the free light-heavy neutrino mixing parameters. If this decay mode dominates the charged Higgs boson decays, it can provide a new way to test the low-scale seesaw mechanism~\cite{Dev:2013wba,Das:2015toa,Khachatryan:2016olu,Das:2016hof,Das:2017zjc,Das:2017gke,Das:2017nvm,Das:2017rsu,Das:2018hph,Bhardwaj:2018lma}. On the other hand, the branching ratio of $H^+ \rightarrow \tau^+\nu$ can also be large for $\tan\beta\simeq \mathcal{O}(50)$. If the decay $H^+ \rightarrow \tau^+\nu$ dominates the charged Higgs boson decays, a lower limit on the charged Higgs boson mass applies with $M_{H^\pm}>80$ GeV~\cite{Abbiendi:2013hk}. Upper limits on $\sigma(pp\rightarrow t b H^+)\times \mathcal{B}(H^+ \rightarrow \tau^+\nu)$ have also been obtained for $M_{H^\pm}=90-2000$ GeV~\cite{Aaboud:2018gjj} and $M_{H^\pm}=180-600$ GeV~\cite{Khachatryan:2015qxa}, respectively. Following the discussions made explicitly in refs.~\cite{Iguro:2017ysu,Iguro:2018qzf,Gori:2017tvg}, which are sufficient for the current purpose, we have found that all these experimental bounds can be satisfied by the parameter regions allowed by the $R_{D^{(\ast)}}$ and $R_{K^{(\ast)}}$ anomalies. As a further nonzero $\epsilon_{ct}$ needs to be introduced to $X_1^u$ in order to provide a resolution to the muon $g-2$ excess while complying with the tight constraints from the $B$-physics observables~\cite{Li:2018aov}, we plan to perform a detailed study of the direct LHC constraints on the charged and neutral scalars at nonzero values of $\epsilon_{tt}$, $\epsilon_{tc}$ and $\epsilon_{ct}$, as well as the neutrino Yukawa couplings in an upcoming paper.

\section{Conclusions}
\label{conclusion}

Based on the structure of the 2HDM-III that has been proposed to address the $R_{D^{(\ast)}}$ anomalies, we have considered a unified scenario where right-handed heavy neutrinos are introduced to the model, so as to generate small neutrino masses and, at the same time, provide reasonable explanation for the $R_{K^{(\ast)}}$ anomalies.

Our main conclusions can be summarized as follows: Within the 2HDM-III, the current world-averaged results for the ratios $R_{D^{(\ast)}}$ can be accommodated at $2\sigma$ level, under the constraint from $\mathcal{B}(B_c^-\rightarrow \tau^- \bar{\nu})\leqslant30\%$. For the light neutrino mass problem, only two massive neutrinos are produced with the sub-eV scale being accounted for by (i) two nearly degenerate Majorana neutrinos with mass around the electroweak scale, (ii) a decoupled heavy Majorana neutrino with mass around $10^{10}~{\rm GeV}$, and (iii) a large $\tan\beta$ with value around $\mathcal{O}(50)$. For the $R_{K^{(\ast)}}$ anomalies, we found that a muon-philic neutrino Yukawa coupling as well as a new top-quark Yukawa coupling, with both of their sizes being of $\mathcal{O}(1)$, are required to reproduce the $1\sigma$ range of the Wilson coefficients in the direction $C_{9 \mu}^{\rm NP}=-C_{10 \mu}^{\rm NP}<0$. Such a large neutrino Yukawa coupling indicates that the coupling in the electron channel should be largely suppressed so as to comply with the constraint from $\mu \rightarrow e \gamma$ while the coupling in the tauonic channel is less constrained from $\tau \rightarrow \mu \gamma$, particularly for heavier charged Higgs boson and right-handed neutrinos.

\section*{Acknowledgements}

This work is supported by the National Natural Science Foundation of China under Grant Nos.~11675061, 11775092 and 11435003. X.L. is also supported in part by the self-determined research funds of CCNU from the colleges' basic research and operation of MOE~(CCNU18TS029). X.Z. is also supported by the China Postdoctoral Science Foundation (2018M632897).

\bibliographystyle{JHEP}
\bibliography{References}

\end{document}